\documentclass[iop]{emulateapj}
\usepackage{soul}
\usepackage{graphicx}
\usepackage{cleveref}
 \usepackage{amssymb}
\usepackage{epstopdf}
\usepackage{apjfonts}
\usepackage[usenames,dvips]{color}
\usepackage{amsmath}
\usepackage{threeparttable}
\usepackage{array,multirow}
\usepackage[normalem]{ulem}

\usepackage{seqsplit}
\makeatletter
\newcommand{\thickhline}{%
    \noalign {\ifnum 0=`}\fi \hrule height 1pt
    \futurelet \reserved@a \@xhline
}
\newcolumntype{"}{@{\hskip\tabcolsep\vrule width 1pt\hskip\tabcolsep}}
\makeatother
\usepackage[bookmarks,breaklinks]{hyperref}
   \hypersetup{
     colorlinks,
     citecolor=blue,
     linkcolor=blue,
    }

\newcommand\ax{1}
\newcommand\bx{2}
\newcommand\cx{3}
\newcommand\dx{11}
\newcommand\ex{12}
\newcommand\fx{6}
\newcommand\gx{4}
\newcommand\hx{5}
\newcommand\ix{25}
\newcommand\jx{7}
\newcommand\kx{8}
\newcommand\lx{9}
\newcommand\mx{10}
\newcommand\nx{19}
\newcommand\ox{13}
\newcommand\px{14}
\newcommand\qx{20}
\newcommand\rx{15}
\newcommand\sx{16}
\newcommand\tx{17}
\newcommand\ux{23}
\newcommand\vx{18}
\newcommand\xx{21}
\newcommand\yx{24}
\newcommand\zx{22}

\usepackage{refcount}% http://ctan.org/pkg/refcount
\newcounter{fncntr}
\newcommand{\fnmark}[1]{\refstepcounter{fncntr}\label{#1}\footnotemark[\getrefnumber{#1}]}
\newcommand{\fntext}[2]{\footnotetext[\getrefnumber{#1}]{#2}}

\begin{document}
\shorttitle{The event-horizon-scale structure of SAGITTARIUS A*}
\shortauthors{Lu et al.}
\title{Detection of intrinsic source structure at $\sim$3 Schwarzschild radii with Millimeter-VLBI observations of SAGITTARIUS A*}
\author{Ru-Sen \ Lu\altaffilmark{\ax,\bx}, 
Thomas P. Krichbaum\altaffilmark{\ax},
Alan L. Roy\altaffilmark{\ax}, 
Vincent L. \ Fish\altaffilmark{\bx}, 
Sheperd S. \ Doeleman\altaffilmark{\bx,\cx}, 
Michael D. \ Johnson\altaffilmark{\cx},  
Kazunori \ Akiyama\altaffilmark{\gx,\hx,\bx},
Dimitrios Psaltis\altaffilmark{\fx}, 
Walter \ Alef\altaffilmark{\ax}, 
Keiichi Asada\altaffilmark{\jx}, 
Christopher Beaudoin\altaffilmark{\bx}, 
Alessandra Bertarini\altaffilmark{\ax,\kx}, 
Lindy Blackburn\altaffilmark{\cx}, 
Ray Blundell\altaffilmark{\cx}, 
Geoffrey C. Bower\altaffilmark{\lx}, 
Christiaan Brinkerink\altaffilmark{\mx}, 
Avery E. \ Broderick\altaffilmark{\dx,\ex}, 
Roger Cappallo\altaffilmark{\bx}, 
Geoffrey B. Crew\altaffilmark{\bx}, 
Jason Dexter\altaffilmark{\ox}, 
Matt Dexter\altaffilmark{\px}, 
 Heino Falcke\altaffilmark{\mx,\ax,\rx}, 
 Robert Freund\altaffilmark{\fx}, 
 Per Friberg\altaffilmark{\sx}, 
 Christopher H. Greer\altaffilmark{\fx},
 Mark A. Gurwell\altaffilmark{\cx}, 
 Paul T. P. Ho\altaffilmark{\jx}, 
 Mareki Honma\altaffilmark{\gx,\tx}, 
 Makoto Inoue\altaffilmark{\jx}, 
 Junhan Kim\altaffilmark{\fx},
 James Lamb\altaffilmark{\vx}, 
 Michael Lindqvist\altaffilmark{\nx},
 David Macmahon\altaffilmark{\px},
 Daniel P. Marrone\altaffilmark{\fx}, 
 Ivan Mart{\'{\i}}-Vidal\altaffilmark{\nx},
 Karl M. Menten\altaffilmark{\ax},
 James M. Moran\altaffilmark{\cx}, 
 Neil M. Nagar\altaffilmark{\qx},
 Richard L. Plambeck\altaffilmark{\px}, 
 Rurik A. Primiani\altaffilmark{\cx}, 
 Alan E. E. Rogers\altaffilmark{\bx}, 
 Eduardo Ros\altaffilmark{\ax,\xx,\zx},
 Helge Rottmann\altaffilmark{\ax}, 
 Jason SooHoo\altaffilmark{\bx}, 
 Justin Spilker\altaffilmark{\fx,\ux},
 Jordan Stone\altaffilmark{\fx},
 Peter Strittmatter\altaffilmark{\fx}, 
 Remo P. J. Tilanus\altaffilmark{\mx,\yx}, 
 Michael Titus\altaffilmark{\bx}, 
 Laura Vertatschitsch\altaffilmark{\cx}, 
 Jan Wagner\altaffilmark{\ax,\ix}, 
 Jonathan Weintroub\altaffilmark{\cx}, 
 Melvyn Wright\altaffilmark{\px}, 
  Ken H. Young\altaffilmark{\cx}, 
 J. Anton Zensus\altaffilmark{\ax}, 
 \& Lucy M. Ziurys\altaffilmark{\fx}
}
\altaffiltext{\ax}{Max-Planck-Institut f{\"u}r Radioastronomie, Auf dem H{\"u}gel 69, D-53121 Bonn, Germany}
\altaffiltext{\bx}{MIT Haystack Observatory, 99 Millstone Road, Westford, MA 01886, USA}
\altaffiltext{\cx}{Harvard-Smithsonian Center for Astrophysics, 60 Garden Street, Cambridge, MA 02138, USA}
\altaffiltext{\gx}{National Astronomy Observatory of Japan, Osawa 2-21-1, Mitaka, Tokyo 181-8588, Japan}
\altaffiltext{\hx}{Department of Astronomy, Graduate School of Science, The University of Tokyo, 7-3-1 Hongo, Bunkyo-ku, Tokyo 113-0033, Japan}
\altaffiltext{\fx}{Steward Observatory and Department of Astronomy, University of Arizona,
933 North Cherry Avenue, Tucson, AZ 85721-0065, USA}
\altaffiltext{\jx}{Academia Sinica Institute of Astronomy and Astrophysics, P.O. Box 23-141, Taipei 10617, Taiwan, ROC}
\altaffiltext{\kx}{Institute of Geodesy and Geoinformation, University of Bonn, D-53113 Bonn, Germany}
\altaffiltext{\lx}{Academia Sinica Institute of Astronomy and Astrophysics, 645 N. A'oh\=ok\=u Place, Hilo, HI 96720, USA}
\altaffiltext{\mx}{Department of Astrophysics/IMAPP, Radboud University Nijmegen, P.O. Box 9010, 6500 GL, Nijmegen, The Netherlands}
\altaffiltext{\dx}{Perimeter Institute for Theoretical Physics, 31 Caroline Street North,Waterloo, ON N2L 2Y5, Canada}
\altaffiltext{\ex}{Department of Physics and Astronomy, University of Waterloo, 200 University Avenue West, Waterloo, ON N2L 3G1, Canada}
\altaffiltext{\ox}{Max Planck Institute for Extraterrestrial Physics, Giessenbachstr. 1, D-85748, Garching, Germany}
\altaffiltext{\px}{University of California Berkeley, Department of Astronomy, Radio Astronomy Laboratory, 501 Campbell, Berkeley, CA 94720-3411, USA}
\altaffiltext{\rx}{ASTRON, The Netherlands Institute for Radio Astronomy, Postbus 2, NL-7990 AA Dwingeloo, The Netherlands}
\altaffiltext{\sx}{James Clerk Maxwell Telescope, East Asia Observatory, 660 N. A'oh\=ok\=u Place, Hilo, HI 96720, USA}
\altaffiltext{\tx}{Graduate University for Advanced Studies, Mitaka, 2-21-1 Osawa, Mitaka, Tokyo 181-8588, Japan}
\altaffiltext{\vx}{Owens Valley Radio Observatory, California Institute of Technology, 100 Leighton Lane, Big Pine, CA 93513-0968, USA}
\altaffiltext{\nx}{Department of Space, Earth and Environment, Chalmers University of Technology, Onsala Space Observatory, SE-439 92 Onsala, Sweden}
\altaffiltext{\qx}{Astronomy Department, Universidad de Concepci\'on, Casilla 160-C, Concepci\'on, Chile}
\altaffiltext{\xx}{Observatori Astron\`omic, Universitat de Val\`encia, C/ Catedr\'atico Jos\'e Beltr\'an no. 2, E-46980 Paterna, Val\`encia, Spain}
\altaffiltext{\zx}{ Departament d'Astronomia i Astrof\'isica, Universitat de Val\`encia, C/ Dr. Moliner 50, E-46100 Burjassot, Val\`encia, Spain}
\altaffiltext{\ux}{Department of Astronomy, University of Texas at Austin, 2500 Speedway, Stop C1400, Austin, TX 78712, USA}
\altaffiltext{\yx}{Leiden Observatory, Leiden University, P.O. Box 9513, 2300 RA Leiden, The Netherlands}
\altaffiltext{\ix}{Korea Astronomy and Space Science Institute, 776 Daedeokdae-ro, Yuseong-gu, Daejeon 305-348, Republic of Korea}

\begin{abstract}
We report results from very long baseline interferometric (VLBI) observations of the supermassive black hole in the Galactic center, Sgr\,A*, at 1.3\,mm (230\,GHz). The observations were performed in 2013 March using six VLBI stations in Hawaii, California, Arizona, and Chile. Compared to earlier
observations, the addition of the APEX telescope in Chile almost doubles the longest baseline length in the array, provides additional {\it uv} coverage in the N--S direction, and leads to a spatial resolution of $\sim30\,\mu$as ($\sim$3 Schwarzschild radii) for Sgr\,A*. The source is detected even at the longest baselines with visibility amplitudes of  $\sim$4--13\,\% of the total flux density. We argue that such flux densities cannot result from interstellar refractive scattering alone, but indicate the presence of compact intrinsic source structure on scales of $\sim$3 Schwarzschild radii. The measured nonzero closure phases rule out point-symmetric emission. We discuss our results in the context of simple geometric models that capture the basic characteristics and brightness distributions of disk- and jet-dominated models and show that both can reproduce the observed data. Common to these models are the brightness asymmetry, the orientation, and characteristic sizes, which are comparable to the expected size of the black hole shadow. Future 1.3\,mm VLBI observations with an expanded array and better sensitivity will allow a more detailed imaging of the horizon-scale structure and bear the potential for a deep insight into the physical processes at the black hole boundary.

\end{abstract}
\keywords{Galaxy: center -- submillimeter: general -- techniques: high angular resolution -- techniques: interferometric}

\section{INTRODUCTION}
\label{sect:introduction}
Most, if not all galaxies, including the Milky Way, are widely believed to harbor supermassive black holes at their 
centers~\citep{1984ARA&A..22..471R,1995ARA&A..33..581K}. It is now widely accepted that the compact source at the center of the Milky Way (Sagittarius A*, hereafter Sgr\,A*) is associated with a $4\times10^6\,M_{\odot}$ supermassive black hole~\citep{2016ApJ...830...17B}, which, due to its proximity ($\sim$8\,kpc), spans the largest angle on the sky among all known black holes~\citep{2001ARA&A..39..309M,2010RvMP...82.3121G,2013CQGra..30x4003F}. For Sgr\,A*, one Schwarzschild radius ($R_{s}$) is $\sim$0.1 au, which subtends an angle of $\sim$10\,$\mu$as to an observer on the Earth. This scale is now within reach with global very long baseline interferometry (VLBI) at a wavelength of 1.3\,mm.

According to general relativity (GR), a lensed image of the accretion disk is punctuated by the black hole silhouette outlined by the image photon orbit around the event horizon of Sgr\,A*~\citep[the latter is known popularly as the ``black hole shadow'';][]{1973blho.conf..215B,1979A&A....75..228L,2000ApJ...528L..13F} and can now be resolved by the Event Horizon Telescope (EHT). This is a project to assemble a VLBI network of millimeter wavelength dishes that aims to resolve general relativistic signatures in the vicinity of nearby supermassive black holes and to generate the first ever black hole image with horizon-scale resolution~\citep{2008Natur.455...78D,2009astro2010S..68D,2011ApJ...727L..36F,2012Sci...338..355D,2015ApJ...807..150A,J2015}. 

Previous VLBI observations at 7 and 3.5\,mm have measured the intrinsic size of Sgr\,A*, providing strong evidence for the existence of a black hole through the implied small emission volume and high density~\citep{2004Sci...304..704B,2005Natur.438...62S,2014ApJ...790....1B}. However, interstellar scattering strongly blurs the image of Sgr\,A* at these wavelengths. Later observations at 1.3\,mm, where the scattering is largely reduced due to the $\lambda^2$-dependence of the angular broadening effect, offered strong evidence that the image of the emitting region has a size comparable to that of the expected black hole shadow for Sgr\,A*~\citep{2008Natur.455...78D,2011ApJ...727L..36F}. Because of the very small number of interferometric baselines used, the early data can only constrain the characteristic size of this image but not any of its detailed properties. 

General relativistic magnetohydrodynamic (GRMHD) simulations of low-luminosity accretion flows around supermassive black holes suggest a number of horizon-scale structures that may be observable in millimeter/submillimeter VLBI images of Sgr\,A*. The images calculated in simulations and semi-analytical models with highly turbulent magnetic fields (the so-called SANE simulations; see~\citealt{Narayan2012}) are often dominated by emission from hot electrons in the accretion flow and generate crescent-like structures~\citep{2009ApJ...697...45B,Dexter2009, Dexter2010, Moscibrodzka2009, 2013A&A...559L...3M, 2015ApJ...799....1C}. When, on the other hand, the emission is dominated by hot electrons in well ordered and strong magnetic fields (e.g., MAD simulations; see~\citealt{McKinney2009,Narayan2012}) then compact and filamentary structures often appear in the simulated images at the footprints of a one- or two-sided jet, possibly showing disjoint emission regions~\citep{2013A&A...559L...3M, 2015ApJ...799....1C}. Similar compact and often disjoined structures also appear in simulated images of orbiting compact emitting regions, or ``hot spots''~\citep{2006MNRAS.367..905B,2012A&A...537A..52E}.  In all of these cases, even though the overall size of the millimeter emission region is comparable to the black hole shadow, i.e., $\sim5 R_{\rm S}$, the images have substantial substructure that is determined by the thermodynamic and magnetohydrodynamic properties of the plasma. Observing such substructure and measuring its properties will provide new insights into the plasma processes, which are acting in the immediate vicinity of a black hole.

In Spring 2013, Sgr\,A* was observed with six VLBI stations in Hawaii, California, Arizona, and Chile at 1.3\,mm. These stations form a subset of the sites that will comprise the planned EHT\fnmark{first-fn}\fntext{first-fn}{The CARMA array telescopes ceased operation in 2015.}. Results obtained from a subset of the 2013 1.3\,mm VLBI data presented in this paper have been published earlier, focusing on the  measurement of high linear polarization~\citep{J2015} at 50--100\,$\mu$as scales ($\geq$ 5--10 $R_{\rm S}$) and on the detection of nonzero closure phases in the Arizona--California--Hawaii triangle~\citep{2016ApJ...820...90F}. In this paper, we reanalyze all 2013 VLBI data with the addition of the  APEX telescope to the array, which allows us to form a more complete VLBI data set with more baselines and closure relations. Our analysis yields the detection of Sgr\,A* on the longest VLBI baselines reported so far (up to 7.3\,G$\lambda$). In the following, we will focus on the small-scale total intensity structure of Sgr\,A* obtained with this extended 1.3\,mm VLBI array, which consists of six stations at four locations (Table~\ref{Table:array}).

\section{Observations, data reduction, and calibration}
\label{sect:obs}
\begin{figure}
\begin{center}
\includegraphics[width=0.5\textwidth,clip]{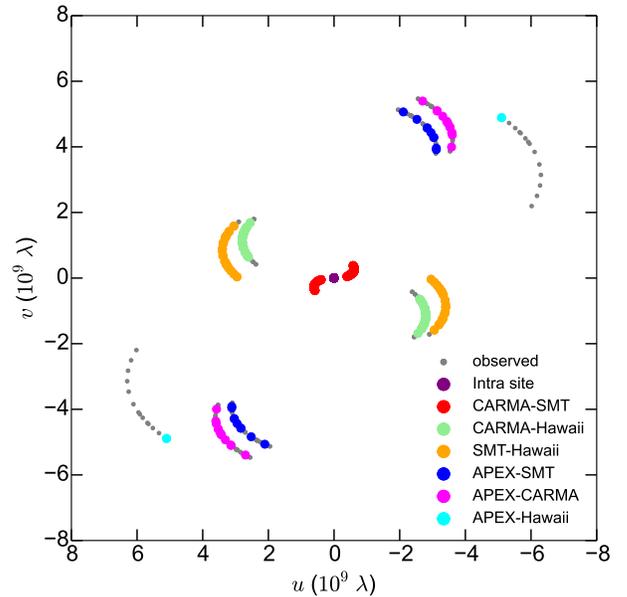}
\caption{Plot of the {\it uv} coverage for the 1.3\,mm VLBI experiments on 
Sgr\,A* in 2013 (gray) with detected scans color-coded by baseline. With regard 
to previous observations, the {\it uv} coverage is significantly improved by 
the addition of the APEX telescope, which adds north--south resolution and 
baselines in the range from 7174\,km (APEX-SMT) to 9447\,km (APEX-Hawaii). Notice that these baselines, for a wavelength of 1.3\,mm, correspond to resolutions of 181\,picoradians and 138\,picoradians, that is, 37.4\,$\mu$as and 28.4\,$\mu$as, respectively.}
\label{fig:uv}
\end{center}
\end{figure}

EHT observations of Sgr\,A* at 230\,GHz were performed on March 21, 22, 23, 26, and 27 (days 80, 81, 82, 85, and 86, respectively) in 2013 with telescopes located at four geographical sites: the Arizona Radio Observatory Submillimeter Telescope (SMT) on Mount Graham in Arizona, USA; the phased Combined Array for Research in Millimeter-wave Astronomy (CARMA) array and one single CARMA comparison antenna in California, USA; The James Clerk Maxwell Telescope (JCMT) and the phased Submillimeter Array (SMA) on Maunakea in Hawaii, USA; and the APEX telescope~\citep{2006A&A...454L..13G,R12,2015A&A...581A..32W} in Chile (see Table~\ref{Table:array}). Figure~\ref{fig:uv} shows the {\it uv} coverage of these observations highlighting those with detected fringes. All sites except the CARMA reference antenna recorded two 480-MHz bands centered on 229.089\,GHz (hereafter low band) and on 229.601\,GHz (hereafter high band), respectively. The CARMA phased array, single CARMA comparison antenna, and SMT simultaneously recorded both left circular polarization (LCP) and right circular polarization (RCP), while the remaining stations (SMA, JCMT, and APEX) recorded a single polarization at a time. Because the quarter-wave plates on each of the SMA antennas can be rapidly rotated, the SMA recorded RCP for 30\,s before each 8 minute long scan, which was then recorded in LCP. These stations, as indicated by one letter codes per polarization used hereafter, are summarized in Table~\ref{Table:array}.

\begin{table*}
\centering
\caption{Array Description\label{Table:array}} 
\begin{tabular}{lllll}
\hline
\hline
Telescope &ID&Polarization&Effective Aperture&Note\\
&&&(m)&\\
\hline
CARMA (single)&D/E&LCP/RCP&10.4&single dish (low band only)\\
CARMA (phased)&F/G&LCP/RCP &25.5/24.1&$5\times10.4\,{\mathrm m} + 3\times6.1\,{\mathrm m}$ 
(day 80); $4\times10.4\,{\mathrm m} + 4\times6.1\,{\mathrm m}$ (days 81--86)\\
JCMT&J&RCP&15.0&JCMT standalone\\
SMA (phased)&P/Q&LCP/RCP&15.9&SMA ($7\times6.0\,{\mathrm m})$; Q for 30\,s scans\\
APEX&A&LCP&12.0&APEX standalone\\
SMT&S/T&LCP/RCP&10.0&SMT standalone\\
\hline
\end{tabular}
\begin{tablenotes}
      \small
      \item Note. The table summarizes telescope names (column 1) and the single letter station code for each polarization (column 2), corresponding polarization of the recorded signals (column 3), effective aperture in meters (column 4), and comments (column 5).
    \end{tablenotes}
\end{table*}

Data were correlated on both the Mark 4 hardware correlator~\citep{2004RaSc...39.1007W} and the DiFX software correlator~\citep[version 2.2,][]{2011PASP..123..275D} at Haystack Observatory. The Mark 4 correlator processed all data except those on baselines to APEX. A recorrelation with the DiFX correlator was performed for all data at times when APEX was observing, but with some disk module failures during this processing. After correlation, the data from the two correlators were merged. 

The data were fringe-fitted using the Haystack Observatory Post-processing System (HOPS) package (version 3.11), which is tailored for millimeter-VLBI data reduction~\citep{1995AJ....109.1391R}. Coherent fringe fitting of all scans was done using the task \texttt{fourfit}. High signal-to-noise ratio (S/N) detections were first used to determine several important quantities for further processing. (1) The phase offsets between the 32 MHz channels within each band were determined. (2) Approximate atmospheric coherence times maximizing the S/N of detection were estimated to guide further incoherent fringe searching. (3) The residual single-band delay, multiband delay, and delay rate were used to set up narrow search windows to lower the fringe detection threshold and to recover low S/N scans. Where possible, the search windows were set using the closure constraints.
After this, data were segmented at time intervals matching the atmospheric coherence time of $4-10$\,s at a grid of values of multiband delay and delay rate, and the amplitudes were time-averaged at all grid cells. A search for a peak in S/N in delay and delay rate space was then performed for each scan to identify the optimal values of delay and rate. This incoherent fringe search lowers the fringe detections threshold in the presence of rapid atmospheric phase fluctuation.
Finally, the detected fringes were segmented at a cadence of 1\,s (which is shorter than the coherence time) and averaged over the scan length to produce noise-debiased estimates of the correlation coefficients based on the incoherent averaging method \citep{1995AJ....109.1391R}.

The correlation coefficients of the Mark 4 data are higher by $\sim$12\,\% than those of the DiFX correlation. An empirical scaling correction was then applied to the Mark 4 correlation coefficients based on the comparison of amplitudes from the two correlations for all available sources and scans.

Closure phases were derived following the same procedure as described in earlier EHT data 
analysis~\citep[e.g.,][]{2012ApJ...757L..14L,2016ApJ...820...90F}. All closure phases were derived based on either the Mark 4 data or DiFX data, but not on the combined data from the two correlators, to avoid nonclosing terms from slightly different correlator models. 
Examination of the fitted residual delays and rates for the detected fringes shows that they close for the available triangles, proving that the closure relations are preserved~\citep{1986A&A...168..365A,1995ASPC...82..189C}.

\subsection{Amplitude calibration}
\label{sect:ampcal}
The a priori calibration of the visibility amplitudes was done using the system equivalent flux density (SEFD) measurements, which were determined by the antenna gain (K Jy$^{-1}$) and the opacity-corrected system temperature. Following previous EHT work~\citep[e.g.,][]{2011ApJ...727L..36F,2012Sci...338..355D,2012ApJ...757L..14L, 2013ApJ...772...13L,2015ApJ...807..150A}, we performed a time-dependent station gain correction on the a priori calibrated amplitudes\fnmark{second-fn}\fntext{second-fn}{Hereafter, we refer to the correction factors from this calibration as gain correction factors, not to be confused with the kelvin to jansky conversion factor.}. 

Our approach has three steps. First, we identified and adopted a total flux density of 3.1\,Jy for Sgr\,A*. This is the average of the reported CARMA local interferometer measurements during the time of VLBI observations, which is consistent with measurements done in parallel at SMA\fnmark{third-fn}\fntext{third-fn}{In measuring the total flux density of Sgr\,A*, the resolution ($\lambda/D_{max}$) of the array for CARMA and SMA is about 3\farcs4 and 3\farcs9, with baselines shorter than 20 k$\lambda$ and 30 k$\lambda$ excluded, respectively.}. The daily average total flux measurements suggest a change at the $\sim$10\,\% level, which is the same order of the a priori calibration accuracy. 

Second, with an intra-site baseline (e.g., DF) and a third station (e.g., S),  one can accurately calculate the gain correction factors for the two co-site stations by assuming that the parallel hand visibilities on the intra-site baselines measure the total flux density of Sgr\,A* and the gain calibrated amplitudes on the other two baselines (e.g., SD/SF) are identical. The lengths of the intra-site baselines (92\,m between the CARMA reference antenna (D/E) and the phased array (F/G) and 156\,m between JCMT and phased SMA) are long enough to resolve out the diffuse thermal emission around Sgr\,A* on arcsecond scales~\citep[with projected baseline length $>$ 26\,m at 1.3\,mm,][]{2016ApJ...820...90F}, but short enough to not resolve the compact source itself at all. We refer to a triangle with such a short baseline as a pseudo closure amplitude triangle~\citep{2015PASJ..tmp..252K}. For the data presented here, we have effectively two pseudo closure amplitude triangles and this calibration can be applied to stations including F/G and D/E (only at low band), and Q and J for the 30\,s scans. 

Third, we transfer the CARMA station gain correction factors obtained at low band to the high band and each 30\,s Q/J correction factors to their following long scan for stations P and J. When transferring the gain correction factors, we have assumed a constant scaling factor during the time of our observations and we set the scaling factors such that the data at both bands from both parallel hands after calibration are self-consistent with each other. Given the known amplitude biases on baselines to station P between the two bands seen in earlier data~\citep{2013ApJ...772...13L}, we used a different P/Q scaling factor for the low and high bands.

We note that the station gain correction factors for SMT and APEX cannot be calibrated in an ``absolute''
manner with the pseudo closure amplitude method. We have assumed that the gain correction factors for SMT and APEX are within 10\,\%--20\,\% from unity. Details of the a priori calibration for APEX can be found in \citet{2015A&A...581A..32W}. We added 10\,\% systematic errors in quadrature to all the calibrated amplitude data to reflect the uncertainties of the a priori gain calibration, although there might still be remaining unaccounted systematic gain offsets for SMT and APEX, which we estimate to be $<$ 20\,\%. Table~\ref{tab:flux} shows the amplitudes of Sgr\,A* after this calibration. 

Our approach differs slightly from the network calibration procedure described in~\citet{J2015} due to, e.g., our improved understanding of the SMT gain that allowed us to directly use the SMT measurements. Nevertheless, our calibrated amplitudes are statistically consistent with the amplitudes reported by \citet{J2015}.

\section{Results}
\label{sect:results}
\subsection{Amplitudes}
\label{sect:amplitude}

\begin{figure*}
\begin{center}
\includegraphics[width=0.485\textwidth,clip]{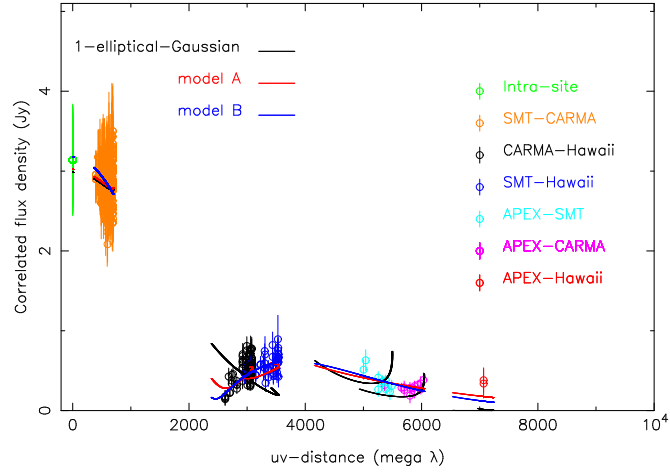}
\caption{Correlated flux density of Sgr\,A* as a function of {\it uv} distance after correcting for the scattering effect (amplitudes on the longest baselines are increased). The fitted amplitudes for the models discussed in Section~\ref{sect:mf} (Table~\ref{tab:model}) are shown in different colors.}
\label{fig:radplt}
\end{center}
\end{figure*}

\begin{figure*}
\begin{center}
\includegraphics[width=0.485\textwidth]{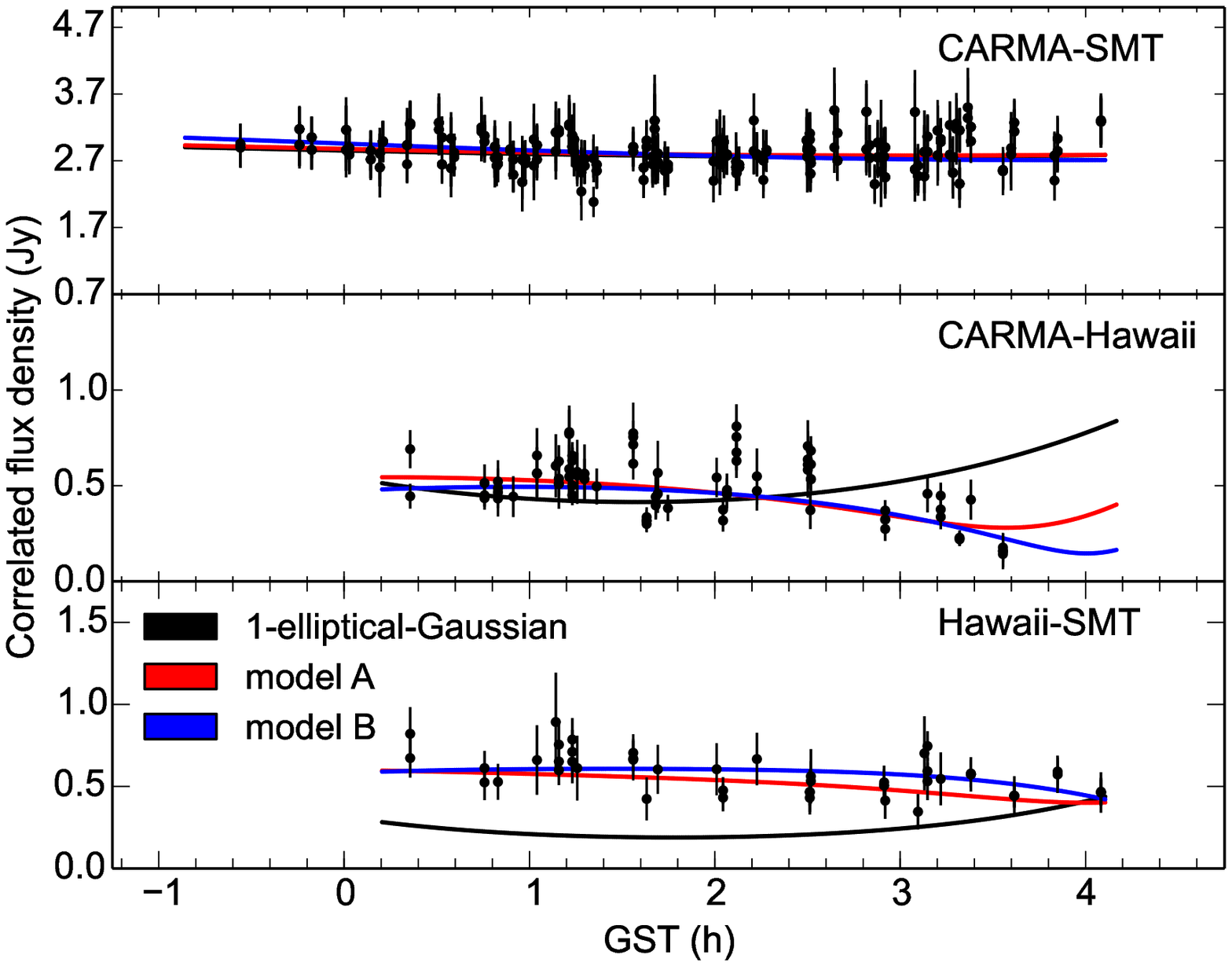}
\includegraphics[width=0.485\textwidth]{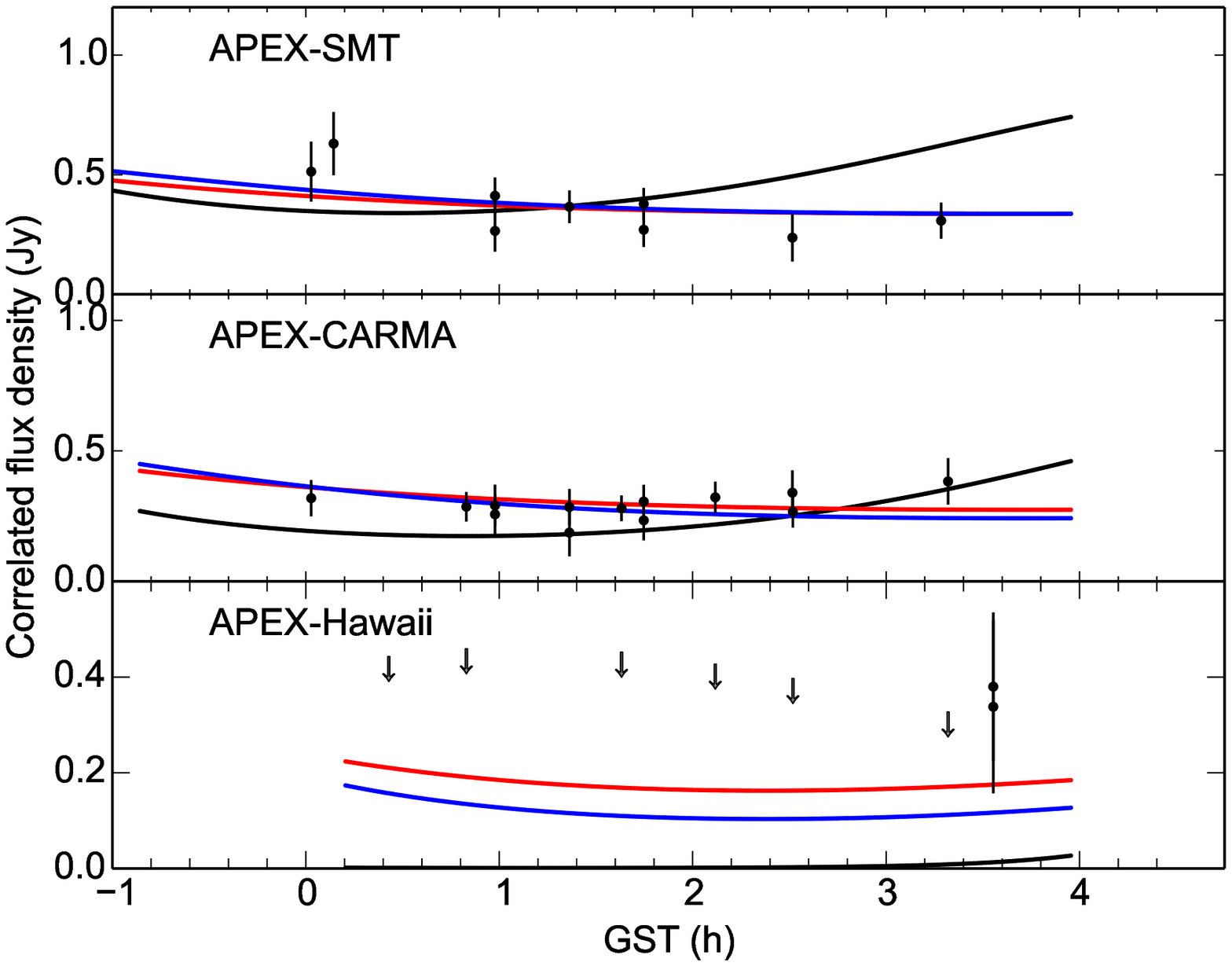}
\caption{Correlated flux density of Sgr\,A* as a function of time (GST) after 
correcting for the scattering effect (error bars correspond to 1$\sigma$). Due to limitations in the amplitude calibration, data points with an ill constrained amplitude calibration are omitted. For the APEX-Hawaii baseline, solid points denote detections and arrows denote 
$5\sigma$ upper limits. The fitted amplitudes for the models discussed in Section~\ref{sect:mf} (Table~\ref{tab:model}) are shown as solid lines in different colors.}
\label{fig:vplt}
\end{center}
\end{figure*}

In the ensemble-averaged limit, the angular broadening of an image due to interstellar scattering
is described by the convolution of the unscattered image with a scattering kernel, 
or equivalently by a multiplication in the Fourier domain. Following ~\citet{2014ApJ...795..134F}, 
we corrected the visibility amplitudes before fitting models of the source structure by 
employing the scattering model determined by~\citet{2006ApJ...648L.127B}. In this model, 
the major axis of the scattering kernel is oriented at $78^\circ$ (east of north), with the associated full width at half maximum (FWHM) for the major and minor axes of
$\theta^\mathrm{maj}_\mathrm{FWHM}$ = 1.309 ($\lambda$/1~cm)$^2$ mas = $22\,\mu$as and 
$\theta^\mathrm {min}_\mathrm{FWHM}$ = 0.64 ($\lambda/$1~cm)$^2$ mas = $11\,\mu$as, 
respectively. The correcting factors, which follow an elliptical Gaussian 
distribution in the Fourier domain, decrease monotonically in all directions 
from unity for the intra-site baselines to $\sim$0.37 for the longest baseline 
between APEX and phased SMA \citep[see ][for the formula for calculating the correcting factors]{Pearson1991}. We show the amplitudes of Sgr\,A* 
after this correction in Figures~\ref{fig:radplt} and \ref{fig:vplt}. 

In Figure~\ref{fig:vplt} (right), we show the amplitudes on the APEX baselines. Due to 
scheduling and technical difficulties, the number of observed VLBI scans to 
APEX was lower than for other stations. On APEX-SMT and APEX-CARMA baselines, Sgr\,A* is detected with S/N in the range of 5--12. However, on the longer APEX-Hawaii 
baseline, Sgr\,A* is detected only in one scan. The S/N of the Sgr\,A* detection 
on this baseline is not very high (5.7 and 7.9 for the low and high band), but 
the low and high bands have very similar delay and delay rates and both are 
close to the values for detections of nearby AGN (e.g., OT\,081 and PKS\,B1921$-$293) scheduled in adjacent scans to Sgr\,A*, which make a false detection 
highly unlikely. A fringe fitting over the combined low and high band data 
gives $\sqrt{2}$ sensitivity improvement and results in a firm detection of 
this scan. In Figure~\ref{fig:vplt}, the upper limits at other times for this 
baseline are also shown. In addition to Sgr\,A*, a few other sources (M\,87, 3C\,273, 3C\,279, 
Centaurus\,A, 4C\,$+$38.41, OT\,081, PKS\,B1921$-$293, and BL Lac) have been robustly 
detected on the APEX to Hawaii baseline during the same observing session. The data for these sources will be presented in forthcoming papers.

\begin{table*}
\centering
\caption[Gain-Corrected Visibility Amplitudes of Sgr\,A*]{Gain-corrected 
Visibility Amplitudes of Sgr\,A*} 
\label{tab:flux}
\resizebox{0.5\textwidth}{!} {
\begin{tabular}{ccccccccc}
\hline
\hline
Day & hh &mm & Baseline & {\it u}& {\it v} & Flux Density & $\sigma$ & Band\\
       &      &       &                & (M$\lambda$)&(M$\lambda$)&(Jy)&(Jy)& (H:high; L:low)\\
\hline      
080& 12&04 &GT &  462.693&   -80.014 &  3.138 &  0.167&  L \\
   & 12&04 &FD &   -0.041&    -0.030 &  3.140 &  0.149&  L \\
   & 12&04 &FS &  462.693&   -80.014 &  2.839 &  0.118&  L \\
   & 12&04 &DS &  462.734&   -79.984 &  2.839 &  0.289&  L \\
   & 12&04 &GE &   -0.041&    -0.030 &  3.140 &  0.237&  L \\
   & 12&04 &ET &  462.734&   -79.984 &  3.138 &  0.365&  L \\
   & 12&05 &GT &  464.476&   -80.997 &  2.771 &  0.030&  L \\   
   & 12 &05 &AS &-3114.326&  3913.996 &  0.342 &  0.051&  L \\
   & 12 &05 &AF &-3578.802&  3994.993 &  0.192 &  0.022&  L \\
   & 15&36 &AP &-5102.190&  4890.027 &  0.142 &  0.021&  L \\
\hline  
\end{tabular}
}
\begin{tablenotes}
\item \textbf{Note.} The times are in UT and the amplitudes are not corrected for blurring (see Section~\ref{sect:amplitude}). $\sigma$ is the measurement uncertainty in flux density.
\item (This table is available in its entirety in machine-readable form.)
\end{tablenotes}
\end{table*}%

\subsection{Closure phases}
\label{sect:cp}
\begin{figure}
\begin{center}
\includegraphics[width=0.38\textwidth,angle=-90,clip,bb=0 0 610 800]{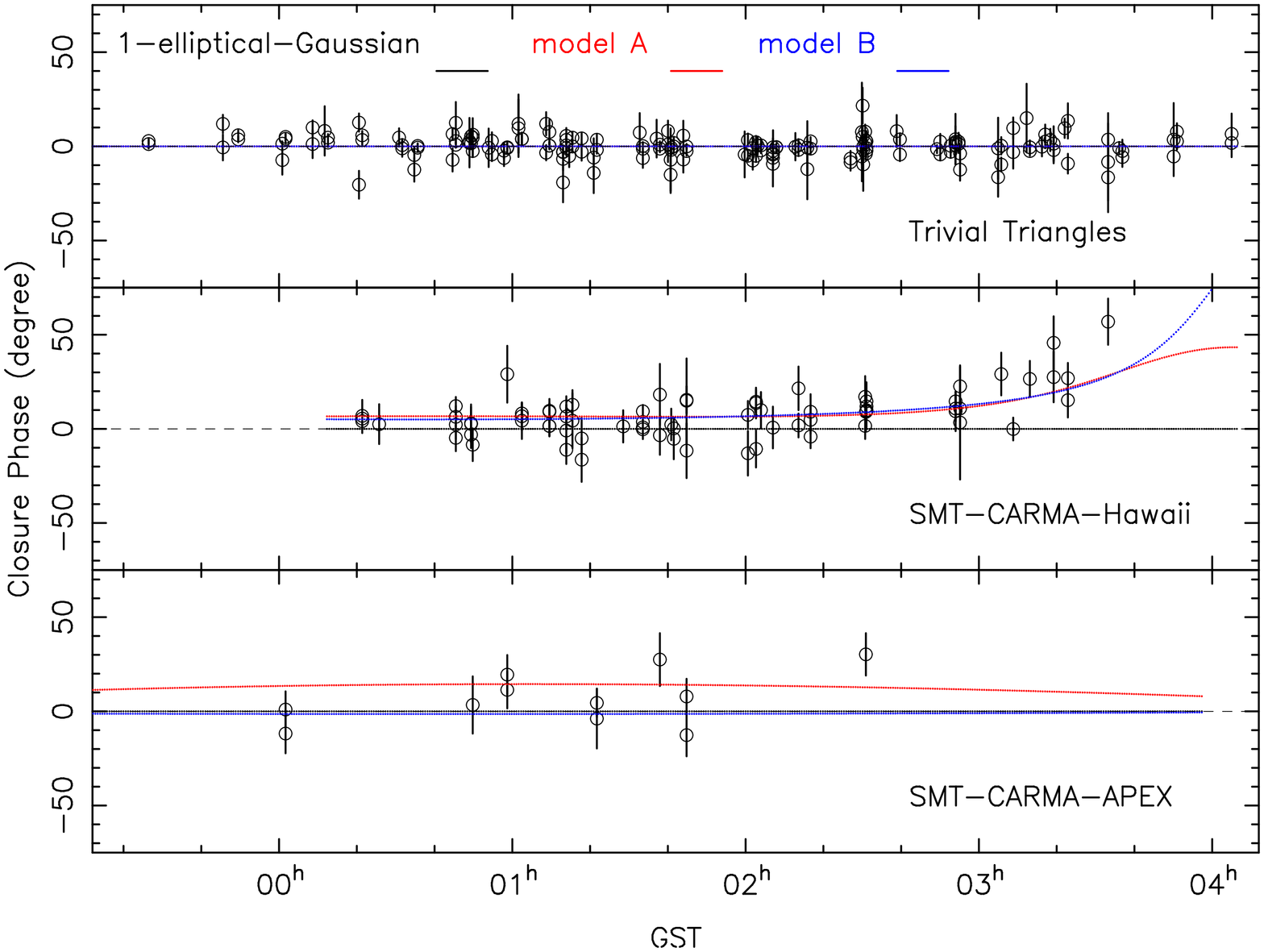}
\caption{Plot of the closure phase of Sgr\,A* as a function of time along with 
the fitted models discussed in Section~\ref{sect:mf} (Table~\ref{tab:model}).}
\label{fig:cphase}
\end{center}
\end{figure}

We show the measured closure phases for Sgr\,A* in Table~\ref{tab:cphase} 
and Figure~\ref{fig:cphase}. The ``trivial'' closure phases, which are formed on 
a triangle including an intra-site baseline, are consistent with zero (median: 
0\fdg2$\pm$0\fdg7 and mean: 0\fdg1$\pm$0\fdg6), as expected, indicating a point-like 
structure at the arcsecond scale resolution provided by the intra-site baselines. On the CARMA--Hawaii--Arizona triangle, we found a median closure phase of 6\fdg7$\pm$1\fdg5  and a mean of 7\fdg8$\pm$1\fdg2 (consistent with an earlier analysis by \citet{2016ApJ...820...90F}), with a trend of increase toward later times during an observing night (Figure~\ref{fig:cphase}). 
Following~\citet{2016ApJ...820...90F}, we can also rule out closure phase errors larger than $\sim$0\fdg2 due to bandpass effects.

We detected Sgr\,A* on the SMT--CARMA--APEX triangle, as well, with an S/N in the range of 4--8. The quality and uncertainties in the measured closure phases for this triangle are not sufficient to quantify their dependence, e.g., on sidereal time. However, we can infer their statistical properties by describing the measurements in terms of a Gaussian mixture model. We find that all 11 closure phase measurements in the SMT--CARMA--APEX triangle are consistent with having the same underlying value of $5.0_{-4.6}^{+12.9}$ degrees, where the uncertainties correspond to a 99.7\% credibility interval ($3\sigma$)\fnmark{fourth-fn}\fntext{fourth-fn}{Unless noted otherwise, all reported confidence intervals are $\pm1\sigma$.}. Compared to the CARMA--Hawaii--Arizona triangle, which is open, small, and oriented mostly along the E--W orientation, the SMT--CARMA--APEX triangle is skinny, larger, and oriented mostly along the N-S orientation. The fact that the closure phases in the SMT--CARMA--APEX triangle are positive and comparable to those measured in the CARMA--Hawaii--Arizona triangle provides additional constraints on the properties of the structure probed by the various baselines of the array we use here, as we will show in the following section.

\begin{table}
\centering
\caption[Closure Phases of Sgr\,A*]{Closure phases of Sgr\,A*}
\label{tab:cphase}
\begin{tabular}{ccccccc}
\hline
\hline
Day & hh & mm& Triangle &Closure Phase &$\sigma$ &Band \\
&&&&(degree)&(degree)&(H:high;L:low)\\
\hline
   080&  12&04 & SDF &  -7.3 &   7.8 & L\\
   &  12&04 & TEG &   1.4 &   1.2 & L\\
   &  12&05 & SFA &   1.0 &   9.8 & L\\
   &  12&05 & SDF &   5.0 &   2.0 & L\\
   &  12&05 & TEG &   4.0 &   2.4 & L\\
   &  12&29 & TGJ &   2.5 &  10.7 & L\\
   &  12&52 & SDF &   2.7 &  12.7 & L\\
   &  12&52 & TEG &   1.8 &  13.4 & L\\
   &  12&53 & DFP &   4.6 &  10.5 & L\\
   &  12&53 & SDF &  -2.5 &   2.8 & L\\
\hline
\end{tabular}
\begin{tablenotes}
\item \textbf{Note.} Times are in UT and $\sigma$ is the measurement uncertainty in the closure phase.
\item (This table is available in its entirety in machine-readable form.)
\end{tablenotes}
\end{table}%

\section{Model Constraints}
\label{section:mc}
In this section, we explore different possibilities for the origin of the observed visibility amplitudes from Sgr\,A* for the long APEX baselines. In particular, we first show that Sgr\,A* cannot be described by a simple symmetric brightness distribution, and that a more complex asymmetric brightness distribution is required.
We then demonstrate that the observed visibility amplitudes at the APEX baselines are too large to be the result of ``refractive noise'' caused by scattering in the interstellar medium but, instead, are indicative of intrinsic substructure in the source image. Finally, we use representative geometric models to argue that our observations are consistent with the  presence of plasma structures, such as crescent images or jet footprints, that are typically generated in GRMHD simulations of low-luminosity accretion flows around black holes.

Hereafter, we will be comparing various geometric models to the visibility amplitude and closure phase data using a Bayesian framework (Akiyama et al. 2018, in preparation; see~\citealt{2009ApJ...697...45B} and~\citealt{2016ApJ...832..156K} for a similar approach on EHT data). If we denote by ``$\mathbf{w}$" the vector of parameters for a given model, then Bayes' theorem allows us to calculate the posterior likelihood $P(\mathbf{w}\vert{\rm data})$ as
\begin{equation}
P(\mathbf{w}\vert{\rm data})=C P_{\rm pr}(\mathbf{w})P({\rm data}\vert\mathbf{w})
\end{equation}
given a prior $P_{\rm pr}(\mathbf{w})$ over the model parameters and a likelihood $P({\rm data}\vert\mathbf{w})$ that the data can be described by this model. $C$ is a normalization constant.

Our data set includes visibility amplitudes measured on intra-site baselines and closure phases on trivial triangles. The latter can be used to test for any biases and inconsistencies in our calibration and pipeline, but do not contribute to the degrees of freedom in a statistical test because all models will predict a zero closure phase for a trivial triangle. Similarly, the intra-site baselines are vital for our amplitude calibration, but do not resolve the source structure. We, therefore, exclude trivial triangles and intra-site baselines from our likelihood calculations.  For the remaining visibility amplitudes and closure phases, we write
\begin{equation}
P({\rm data}\vert \mathbf{w})=\prod_{i=1}^{B+C}\prod_{j=1}^{M}P_{ij}({\rm data}\vert\mathbf{w})\;,
\label{eq:likely}
\end{equation}
where $i=1,...,B+C$ and $j=1,...,M$ denote baselines/triangles and time instances, respectively, $B$ is the total number of baselines, $C$ is the total number of triangles, and $M$ is the total number of time instances. $P_{ij}({\rm data}\vert\mathbf{w})$ denotes the likelihood that a single measurement in a given baseline and at a given time instance is consistent with the model predictions.

In writing the last equation, we have implicitly assumed that all data points are statistically independent from each other. Formally, this is not appropriate for our data because, e.g., uncertainties in the gains of individual telescopes lead to covariant uncertainties in the model visibility amplitudes. In principle, given that we do not have a perfect knowledge of the telescope gains, we would write the likelihood by marginalizing over all possible values of the gains, i.e.,
\begin{equation}
P({\rm data}\vert \mathbf{w})=\prod_{i=1}^{B+C}\prod_{j=1}^{M}\int P_{ij}({\rm data}\vert\mathbf{w},G_{\{k\},j})P(G_{\{k\},j})dG_{\{k\},j}\;,
\label{eq:likely_gains}
\end{equation}
where $G_{\{k\},j}$ is the set of complex gains of $\{k\}$ telescopes at the $j-$th time instance and $P(G_{\{k\},j})$ measures the likelihood of a given set of telescope gains at a given instance in time. The latter is our model of systematic uncertainties in the telescope gains, which results from our amplitude calibration (Section~\ref{sect:ampcal}). If the uncertainties in the data are well described by a Gaussian distribution (which is true for interferometric data only in the limit of large S/Ns), the likelihoods of the telescope gains are also Gaussian, and no pair of baselines shares a telescope, then equation~(\ref{eq:likely_gains}) is equivalent to equation~(\ref{eq:likely}) with the uncertainties in the measurements and the uncertainties in the gains added in quadrature.

The data we report here have too limited {\it uv} coverage to allow for a detailed comparison with complex models. Instead, our goal below is simply to demonstrate that the simple geometric structures predicted by GRMHD models are consistent with the measured visibility amplitudes and closure phases for realistic values of the model parameters. For these reasons, employing equation~(\ref{eq:likely_gains}) in its full complexity is not warranted for the purposes of the present work. 

Here, we assume, for simplicity, that all data points are uncorrelated and that the remaining gain uncertainties can be added in quadrature to the measurement uncertainties. We then use equation~(\ref{eq:likely}) with an Exchange Monte Carlo (EMC) method~\citep{1996JPSJ...65.1604H}, which is a subclass of the Markov Chain Monte Carlo (MCMC) methods, to calculate the posterior distribution over the parameters of the various models we consider below. For the prior distribution of each parameter, we adopt a uniform distribution. Therefore, the posterior distribution and the likelihood are proportional to each other, leading to the same estimates of best-fit parameters as those one would have obtained from maximum likelihood methods.
Because of the approximations we discussed above, our approach will allow us to identify plausible values for the model parameters that are consistent with the data, but not to fully explore the uncertainties in the model parameters or their covariances.

\begin{figure*}
\begin{center}
\includegraphics[width=0.4\textwidth]{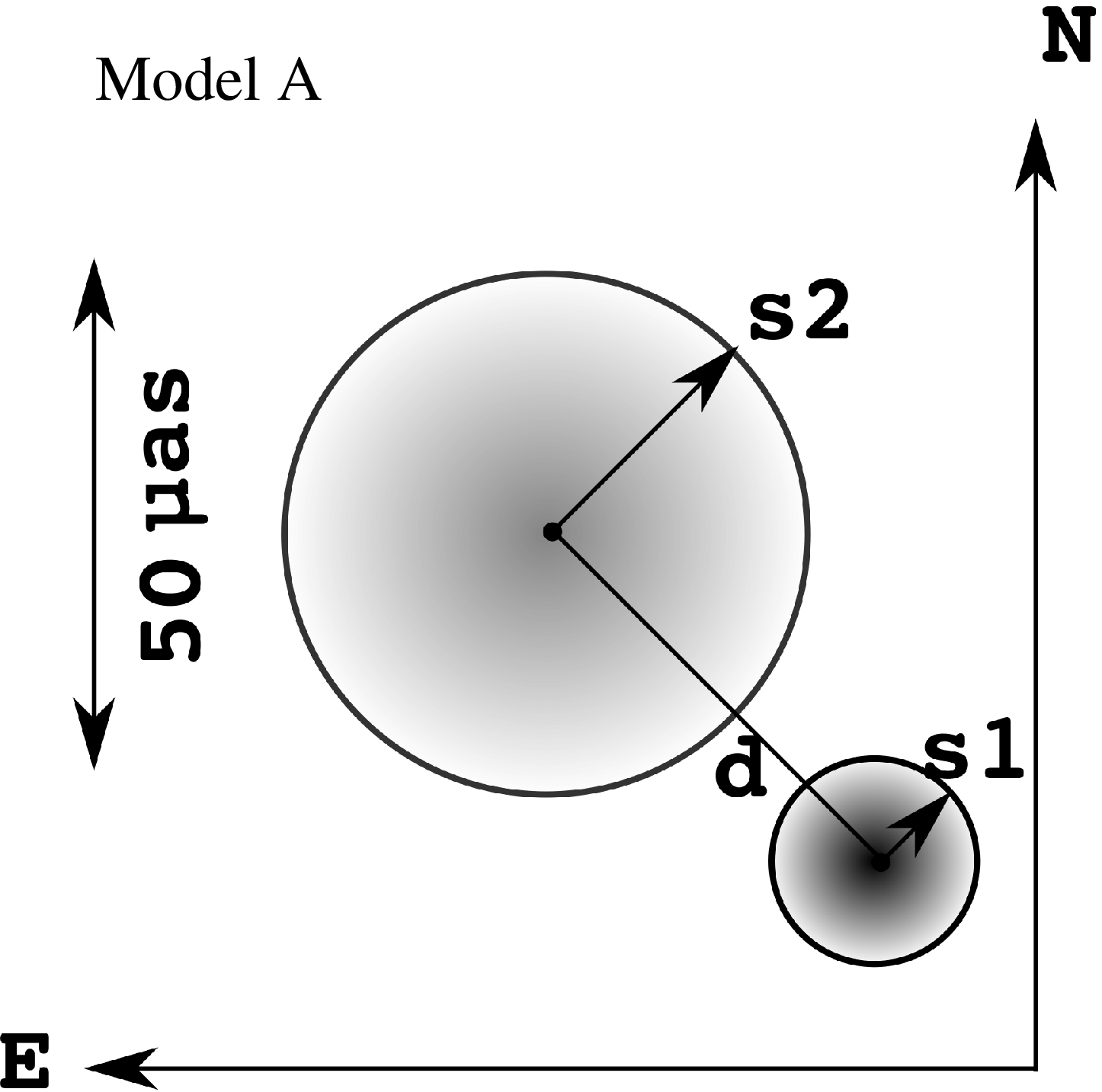}
\includegraphics[width=0.4\textwidth]{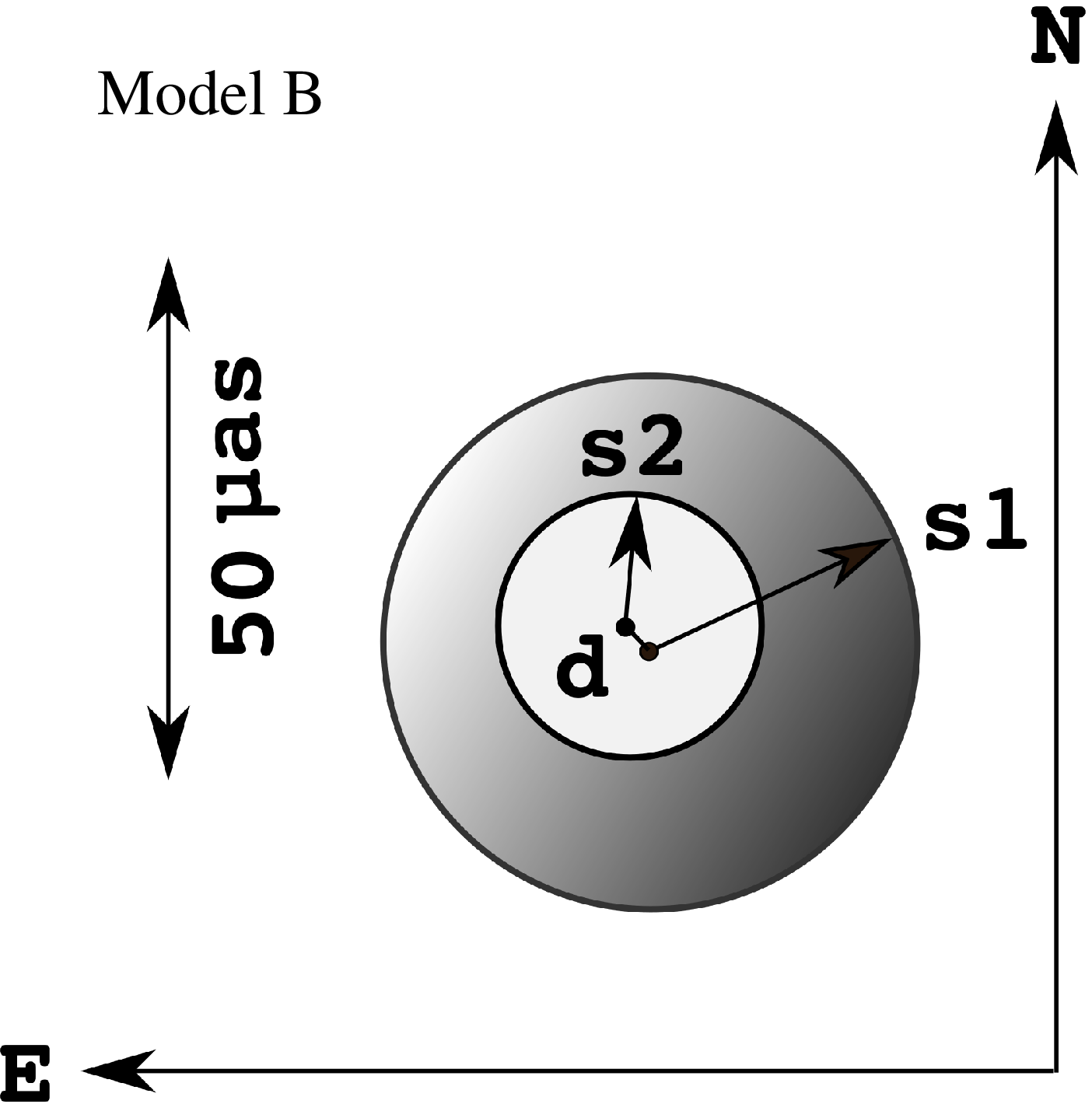}\\
\includegraphics[width=0.45\textwidth]{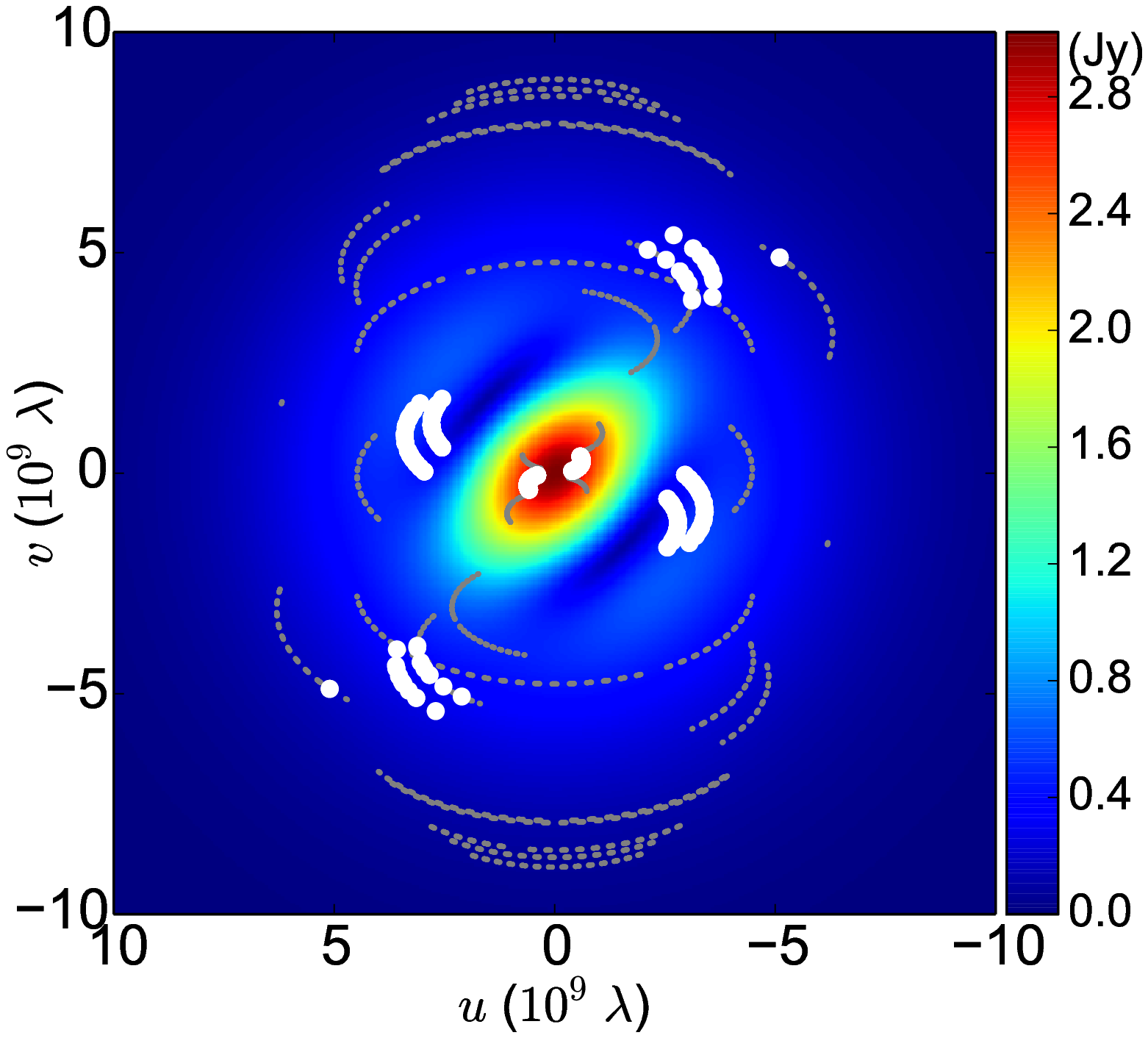}
\includegraphics[width=0.45\textwidth]{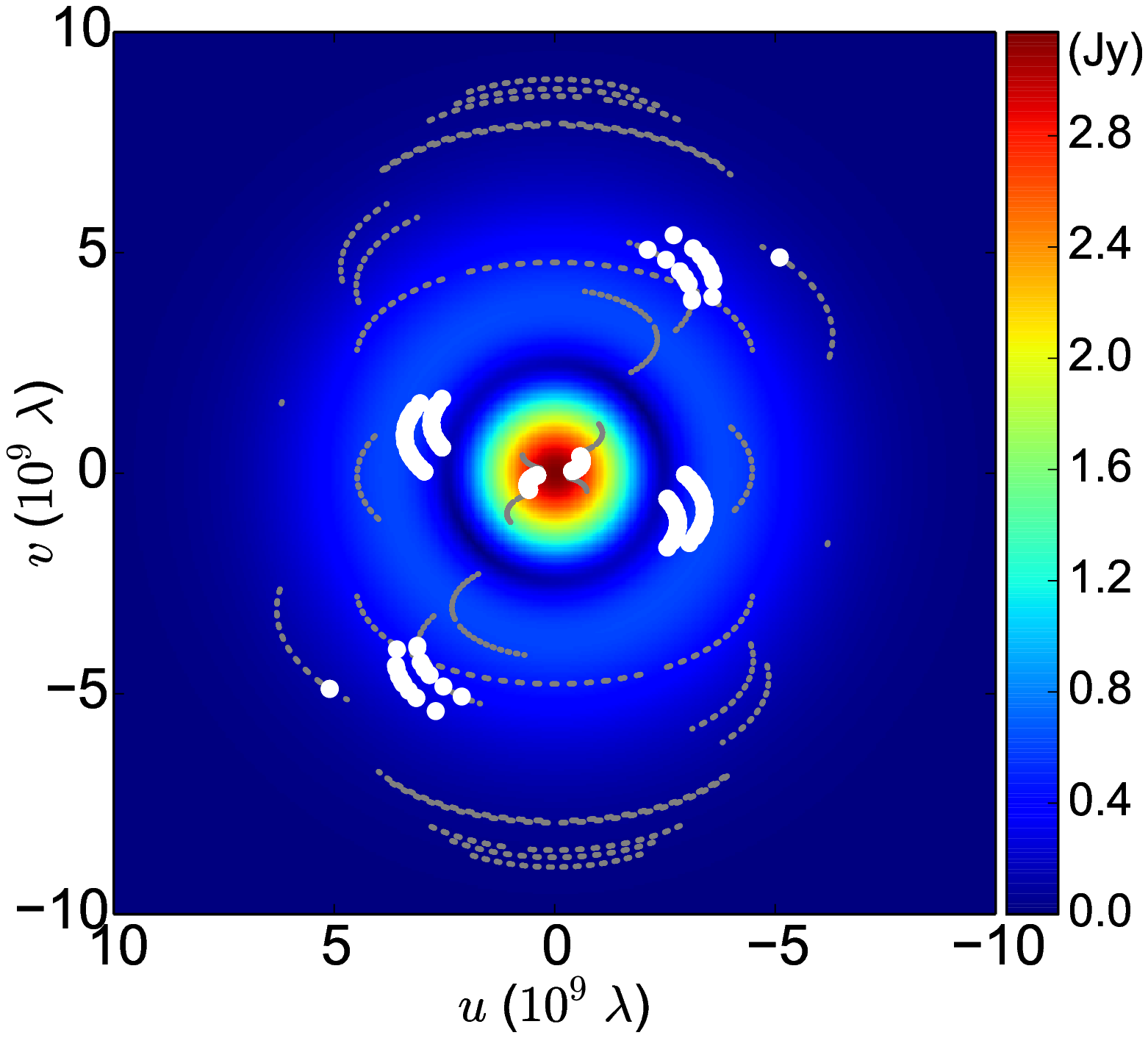}
\caption{ (Top) Schematic representations of the two geometric models (A and B) that we use to assess whether simple image structures can reproduce the observed visibility
amplitudes and closure phases. The relative sizes and orientations of the two components in each model correspond to the most likely values shown in Table~\ref{tab:model} ($s_1$ and $s_2$ indicate half of the FWHM of the corresponding component). The $\simeq$50\,$\mu$as scale of the schematic is shown by the double arrow. (Bottom) Direct Fourier transform {\it uv} maps of the visibility amplitudes for models A (left) and B (right) for the most likely values of their parameters. The gray curves show the $u-v$ tracks of future full-array EHT observations of Sgr\,A*. The white dots show the locations of the measurements reported here.}
\label{fig:image}
\end{center}
\end{figure*}

\setlength\extrarowheight{4pt}
\begin{table*}
\caption{Model-fitting results.}
\centering
\begin{threeparttable}
\begin{tabular}{crrrrcccc}
\hline
\hline
\multirow{2}{*}{}&Flux Density\tnote{a} & $x$\tnote{b}            &             $y$\tnote{b} &Size\tnote{c}&Ratio&P.A.\tnote{d}&\multirow{2}{*}{$k$\tnote{e}}&\multirow{2}{*}{$\Delta$BIC\tnote{f}}\\
                                  &(Jy)             &($\mu$as)&($\mu$as)&($\mu$as)&&(degree)&\\
                                  %\cline{2-9}
                                  \hline
        Single elliptical Gaussian&$3.0\pm0.2$ &0&$0$ & $52\pm1$&  $0.4\pm0.1$         &$81\pm3$&4&980\\           
                                  \hline
\multirow{2}{*}{Model A~}(S1)&$0.9\pm0.1$ &$0$&$0$&$20\pm1$&\multirow{2}{*}{...}&\multirow{2}{*}{...}&\multirow{2}{*}{6}&\multirow{2}{*}{14}\\ 
\multirow{2}{*}{\hspace{30pt}}(S2) &$2.1\pm0.2$ &$31\pm1$&$31\pm3$&$51\pm2$&\\                                          
\hline

\multirow{2}{*}{Model B~}(S1)&$4.9\pm0.6$ &$0$&$0$&$52\pm2$&\multirow{2}{*}{...}&\multirow{2}{*}{...}&\multirow{2}{*}{6}&\multirow{2}{*}{0}\\
\multirow{2}{*}{\hspace{30pt}}(S2) &$-1.7\pm0.4$ &$1\pm1$&$2\pm1$&$25\pm2$&\\      
\hline
\end{tabular}
\begin{tablenotes}
    \item[a] Flux density for each component.
    \item[b] Relative position of each component in R.A. and decl.. The total displacement of the two components is $d=(x^2+y^2)^{1/2}$.
    \item[c] The full width at half maximum (FWHM) of each Gaussian component. 
    \item[d] The position angle of the major axis for the elliptical Gaussian model in degrees east of north.
    \item[e] Number of model parameters.
    \item[f] $\Delta$BIC = BIC-682.
  \end{tablenotes}
\end{threeparttable}
\label{tab:model}
\end{table*}%

\subsection{Stationary resolved structure?}
In principle, the VLBI structure of Sgr\,A* may be variable on short time scales
\citep{2011ApJ...727L..36F,2016ApJ...817..173L, 2016arXiv160106799M,2016A&A...587A..37R}. However, we note that the {\it uv} coverage on the separate observing days is insufficient for a detailed imaging/modeling on a per day basis. We therefore, combined all data into a single data set based on the assumption that the VLBI structure does not change significantly during our observing campaign. This assumption is supported by an unpaired t-test, which shows an insignificant difference in closure phases between the different observing nights \fnmark{fifth-fn}\fntext{fifth-fn}{two-tailed $P$-values ranging from 0.12 to 0.93}. Additionally, we also calculated the compactness ratio R=$S_{\rm long}/S_{\rm short}$ of the average correlated flux density on long (2.9 -- 3.1 G$\lambda$) and short baselines (400 -- 700 M$\lambda$) on a per day basis. We then performed a $\chi^2$-test to check for possible variability of that ratio. The resulting $p=0.81$ excludes that R varies significantly. This is consistent with earlier observations of \citet{2011ApJ...727L..36F}, who showed stationarity of the source size over time-scales of a few days, despite total flux density variations at a level of $\leq 20$\,\%, the latter being of the same order as the amplitude calibration accuracy and the level of total flux density variability observed in this experiment. Owing to the limited number of detections on each observing day, we therefore cannot make a strong statement about a possible underlying structural variability of Sgr\,A*. New data with better and more uniform {\it uv} coverage will be needed for such a statement.

Earlier 1.3\,mm VLBI observations of Sgr\,A* with baselines primarily oriented in the E--W orientation
justified no more than a circular Gaussian model fit, which results in an approximate source size of $37^{+16}_{-10}\,\mu$as (3$\sigma$ errors) for the brightness distribution of Sgr\,A*~\citep{2008Natur.455...78D,2011ApJ...727L..36F}. The addition of the APEX telescope to the VLBI array allows us to constrain the source size also in the N--S direction and almost doubles the angular resolution. In fact, the measurement of a nonzero closure phase for the triangles SMT--CARMA--Hawaii and CARMA--SMT--APEX (both Figure~\ref{fig:cphase} and \citealt{2016ApJ...820...90F}) and the observed visibility amplitudes (Figure~\ref{fig:radplt}) are both inconsistent with a simple elliptical Gaussian model, which is characterized by the different size in the two orthogonal directions (see also~\citealt{J2015}).

In order to demonstrate this, we show in Table~\ref{tab:model} the results for a model fit with one elliptical Gaussian. The inferred size of the major axis is $52\pm 1~\mu$as, the axial ratio is $0.4\pm 0.1$, and the position angle of the major axis is $81^\circ\pm3^\circ$ east of north. However, this model provides a very poor fit to the data, as measured by the difference in the Bayesian information criterion \citep[BIC,][]{2007MNRAS.377L..74L} for this model, in comparison with the more asymmetric models, which are described in \S\ref{sect:mf} (see Table~\ref{tab:model}). For the purposes of our work, we compute the BIC as
\begin{equation}
\mbox{BIC}=k \ln(N) -2 \ln P({\rm data}|\mathbf{w})\;,
\end{equation}
where $k$ and $N$ are the number of model parameters and data points, respectively. Of course, the measurement of a nonzero closure phases is also inconsistent with other point-symmetric models, like a ring or a symmetric annulus. It requires more complex, asymmetric structures.

It is important to emphasize here that the above result does not depend on the precise model of scattering induced blurring that we use. In deblurring the visibilities, we have assumed a Gaussian  kernel at 1.3\,mm with major and minor axis sizes extrapolated from the longer-wavelength measurements following a $\lambda^2$ law. The extrapolated sizes using the slightly different inferences by \citet{2006ApJ...648L.127B} and \citet{2005Natur.438...62S} differ by only $1\,\mu$as at 1.3\,mm, which has little impact on our results. The scattering correction does depend weakly on the assumption of a Gaussian kernel, of isotropy,  and of the quadratic $\lambda^2$ slope of the scattering law. There is evidence that, at millimeter-wavelenghts, the diffractive scale becomes comparable to the inner scale of turbulence and, hence, the angular broadening scaling becomes steeper than $\lambda^2$ and that the shape of the kernel becomes non-Gaussian on long baselines \citep{2014ApJ...794L..14G,2015ApJ...805..180J}. 
To assess the impact of these effects, we deblurred the calibrated data by considering extreme cases of reasonable inner scales of turbulence \citep{2016ApJ...833...74J} and found that the model parameters do not change significantly.

\subsection{Refractive Noise}

In principle, the visibility amplitudes that we have measured on the various APEX baselines may not be caused by intrinsic source structure alone but rather be affected by refractive scattering, which is caused by the interstellar medium, if the APEX baselines already would resolve the underlying source structure. Refractive scattering effects on long VLBI baselines have been detected in Sgr\,A* at 1.3\,cm by \citet{2014ApJ...794L..14G}. The properties of refractive substructure have been calculated by, e.g., \citet{2015ApJ...805..180J} and \citet{2016ApJ...826..170J}. In this subsection, we follow the work of \citet{2016ApJ...826..170J} to demonstrate that the visibility amplitudes observed at the APEX baselines are unlikely to have been caused by refractive scattering effects, but they must represent (at least partially) evidence for intrinsic source substructure. 

For an isotropic scattering medium, the amplitude of refractive noise  (in units of the zero-baseline flux density) at a baseline that resolves the image is given by Equation~(18) in \citet{2016ApJ...826..170J}---see also Figure~7 and Equation~(19) of \citet{2015ApJ...805..180J}---i.e.,
\begin{eqnarray}
\sigma(b)&\simeq&\sqrt{\frac{\Gamma(4/\alpha)\Gamma(1+\alpha/2)}
	{2^{2-\alpha}\Gamma(1-\alpha/2)}}
    \left(\frac{r_0}{r_F}\right)^{2-\alpha}
    \left[\frac{b}{(1+M)r_0}\right]^{-\alpha/2}
    \nonumber\\
&&\qquad \times \left(\frac{\theta_{\rm scat}}{\theta_{\rm img}}\right)^2\;,
\end{eqnarray}
where $\Gamma(x)$ is the gamma function, $\alpha$ is the power-law index of the turbulent power spectrum, $r_0$ is the phase decoherence length, $r_{\rm F}$ is the Fresnel scale, $M\equiv D/R$ is the magnification factor, $D$ is the observer-screen distance, $R$ is the screen-source distance, 
\begin{equation}
\theta_{\rm scat}\equiv \frac{\sqrt{2\ln2}}{\pi} \frac{\lambda}{(1+M)r_0}
\label{eq:JN18}
\end{equation}
is the scattered angular size of a point source, and $\theta_{\rm img}=\sqrt{\theta_{\rm src}^2+\theta_{\rm scat}^2}$ is the ensemble-average angular size of the source.

At 1.3\,mm, $\theta_{\rm scat}\simeq 22~\mu$as~\citep{2006ApJ...648L.127B}, whereas the present observations constrain the size of a potentially resolved source to $\theta_{\rm src}\simeq 52~\mu$as (see Table~\ref{tab:model}). For a screen at 2.7 kpc (as inferred by observations of the galactic center magnetar; \citealt{2015ApJ...798..120B}), the Fresnel scale is equal to $\sim10^{10}$~cm. The APEX baselines have a length of $\simeq$6\,G$\lambda\simeq 8\times 10^8$~cm. Finally, under the hypothesis (which we try to reject) that the correlated flux in the APEX baselines is primarily due to refractive noise, these baselines have resolved the underlying image, i.e., $b>(1+M)r_0$. Using this inequality and substituting the above values into equation (5), we get
\begin{eqnarray}
\sigma(b)&\le&0.065\sqrt{\frac{\Gamma(4/\alpha)\Gamma(1+\alpha/2)}
	{2^{2-\alpha}\Gamma(1-\alpha/2)}}
    \left(\frac{b}{8\times 10^8\,{\rm cm}}\right)^{2-\alpha}
    \nonumber\\
&&\times \left(\frac{r_F}{10^{10}\,{\rm cm}}\right)^{\alpha-2}
\left(\frac{\theta_{\rm scat}}{22~\mu{\rm as}}\right)^2
\left(\frac{\theta_{\rm img}}{56~\mu{\rm as}}\right)^{-2}\;.
\end{eqnarray}
For a 3D Kolmogorov spectrum of turbulence, $\alpha=5/3$ and $\sigma< 2$\%. This value depends weakly on $\alpha$, since for $\alpha=3/2$, $\sigma< 3$\% and for $\alpha=1.9$, $\sigma<1.5$\%. Note that these upper limits are very conservative since we have assumed that the APEX baselines have barely resolved the ensemble-average image of the source.

The visibility amplitudes that we detected on the APEX baselines is $\sim$4--13\,\% of the zero-baseline amplitude (see Table~\ref{tab:flux}), while the dispersion of refractive noise at the same baselines is at most 2-3\,\%. Furthermore, for anisotropic scattering, the amplitude of refractive noise along the N--S direction (i.e., along the minor axis of the scattering kernel) is further suppressed by factors of a few (see, e.g., discussion in \citealt{2016ApJ...826..170J}), making the estimate of the amplitude of the refractive noise for the APEX baselines to be at most 1\,\%. Even though we have only measured one particular realization of the refractive noise over a narrow range of baselines, it is unlikely that we detected a distribution of amplitudes in these baselines at the 4-13\% level caused by refractive scattering effects that are expected to be at the $\sim1$\% level. In other words, it is very unlikely that the visibility amplitudes that we measured in the APEX baselines have resulted predominantly from refractive noise, with no contribution from intrinsic source substructure.

\subsection{Physically Motivated Models}
\label{sect:mf}
 In the previous subsections, we argued that our measurements of the correlated flux on the APEX baselines are indicative of source intrinsic emission on spatial scales comparable to $\sim3$~Schwarzschild radii. We now assess whether physically plausible structures are consistent with the data we report here. 

The sparse {\it uv} coverage of our data does not warrant the reconstruction of a VLBI image of the inner accretion flow or fitting the parameters of detailed GRMHD models to the data. However, as discussed in Section~\ref{sect:introduction}, models of radiatively inefficient accretion flows predict images that are often shaped either like crescents, for disk-dominated models, or like disjoined compact emission regions at the jet footprints, for jet-dominated models. For this reason, we employ here two simple geometric models that were constructed in the past to capture the basic structural characteristics of images that are generated by complex GRMHD simulations (see, e.g., \cite{2013MNRAS.434..765K, 2016arXiv160900055B, 2016arXiv160106799M} and references therein for a discussion). 

Model~A includes two displaced, positive, circular Gaussian components. It provides a simplified generic description of the expected image of the footprints of a jet~\citep[e.g.,][]{2015ApJ...799....1C}, 
although a jet may appear more complex on event horizon scales than described by our Model A
~\citep[e.g.,][]{2014A&A...570A...7M}. It may also be used to describe the dominant emission from a tilted disk~\citep{2013MNRAS.432.2252D} or a compact emission region, which is located off-axis to a more extended emission region, e.g., a hot spot in an accretion flow~\citep{2006MNRAS.367..905B}. Because we cannot measure absolute phases in millimeter-VLBI, our data are only sensitive to relative positions. For this reason, we fix the center of one of the Gaussian components to the origin. Model A then requires six parameters: the flux density of each component, the R.A. and decl. displacement of the second Gaussian component, and the FWHM sizes of each component.

Model~B is meant to provide a simplified generic description of the expected crescent-like emission around a BH for disk-dominated emission models. It is similar to the models of \cite{2013MNRAS.434..765K} and \cite{2016arXiv160900055B} in the sense that it is constructed as a difference between two displaced Gaussian components. The Gaussian taper of Model~B produces smoother variations of the closure phases  in comparison to a similar model consisting of uniform disks~\citep{2013MNRAS.434..765K}. The sharp edges of the latter lead to steep gradients in the closure phase on SMT--CARMA--Hawaii, which are not observed. On the other hand, our model B has fewer parameters than the more complex geometric model of \cite{2016arXiv160900055B}, as warranted by the limited {\it uv} coverage of our current data. Since Model B also involves two Gaussian components (albeit one with a negative normalization), it has the same number of parameters as Model A.

Table~\ref{tab:model} summarizes the results from the model fitting. The single Gaussian model does not fit the data\fnmark{sixth-fn}\fntext{sixth-fn}{$\Delta$BIC=980 for all data, $\Delta$ BIC=918 if closure phases on the SMT--CARMA--Hawaii triangle after 03h GST are excluded for analysis.}.
For the two-component models A and B, we list the most likely values of their parameters 
(uncertainties reflecting formal errors) and the relative difference between their BIC. We note that the difference in BIC between models A and B depends largely on the treatment of the measurement uncertainties and the gain calibration. Although model B formally fits the data better, we cannot rule out model A, owing to the residual uncertainties in the error budget.

Figure~\ref{fig:image} shows the relative sizes and orientations of the model components, the resulting $u-v$ maps, and the locations on the maps where our measurements reside. It is important to emphasize here that we have assumed flat priors for all model distributions with no constraints on the space of model parameters. It is, therefore, instructive that the most likely characteristic sizes of both models, as measured by the component separation in Model~A or the size of the larger component in Model~B, are comparable to 50 $\mu$as, i.e., the expected size of the black hole shadow. This is consistent with the predictions of complex GRMHD simulations: either two jet footprints surrounding the black hole shadow for jet-dominated images or a crescent with a size comparable to that of the shadow for disk-dominated images (see references above). Moreover, in both models, the visibility amplitudes at the APEX baselines are generated by physical structures (i.e., jet footprint sizes or crescent widths) that are smaller than the size of the shadow. 

\section{Summary and Conclusions}
\label{sect:summary}

In this paper, we presented results based on 1.3\,mm VLBI observations of Sgr\,A* with the EHT in 2013. With respect to earlier observations, the baseline coverage is significantly improved by the addition of the APEX telescope, which provides a resolution of $\sim3 R_{\rm S}$ on Sgr\,A* in the N--S direction. The small-scale VLBI structure of Sgr\,A* is resolved and appears asymmetric (nonzero closure phase), as suggested earlier by ~\citet{2016ApJ...820...90F}. We show that the visibilities can be fitted by two Gaussian components of different size and displacement. The marginal difference in the fit quality of the two models cannot yet be distinguished, although model B formally fits the data slightly better.

The measured relatively large visibility amplitudes on the APEX baselines of $\sim$4--13\,\% of the total flux density are not consistent with image substructure solely caused by refractive scattering effects in the interstellar medium. On the other hand, our data can be well fit by geometric images of different morphologies (jets or disks) that capture the basic structure predicted by physically motivated GRMHD simulations of radiatively inefficient accretion flows. Although the limited {\it uv} coverage of our data do not allow us to draw detailed conclusions about the properties of the observed structures, it is instructive that the best-fit models have common structural properties and physically plausible values of parameters, i.e., their brightness asymmetry, their northeast-southwest orientation, and the characteristic sizes of the structures that are comparable to the expected size the black hole shadow. As Figure~\ref{fig:image} shows, the different types of structures that we consider here (jet versus disk) make widely different predictions for the other VLBI baselines that were sampled in recent (April 2017) VLBI observations with a larger array and including ALMA as a new VLBI station. New imaging algorithms, statistical tools, and scattering models are being developed to harness the potential of these new EHT observations, which are expected to provide sufficient {\it uv} coverage to distinguish between different models, allow full imaging of these horizon-scale structures, and provide a new window into physical processes at the black hole boundary.

\acknowledgments
We thank the anonymous referee for constructive comments that improved the paper. The authors thank the APEX observatory staff and the APEX team at the MPIfR for 
their support. In particular, we appreciate the help of R. G\"usten, D. Muders, and A. Weiss.
R.-S. L  thanks Dr. Lei Chen for fruitful discussion on the data analysis and Dr. R. W. Porcas and Dr. N. R. MacDonald for very useful comments and discussions. K.A thanks Prof. Koji Hukushima, Prof. Masato Okada, and Dr. Kenji Nagata for many useful advice and suggestions on the development of our modeling code using the Exchange Monte Carlo method adopted in this work. H.F. acknowledges funding from the European Research Council (ERC) Synergy Grant ``BlackHoleCam'' (Grant 610058). E.R. acknowledges support from the Spanish MINECO through grants AYA2012-38491-C02-01 and AYA2015-63939-C2-2-P and from the Generalitat Valenciana grant PROMETEOII/2014/057. Based on observations with the Atacama Pathfinder EXperiment (APEX) telescope (under project ID 091.F-9312(A)). APEX is a collaboration between the Max Planck Institute for Radio Astronomy, the European Southern Observatory, and the Onsala Space Observatory. The Submillimeter Array is a joint project between the Smithsonian Astrophysical Observatory and the Academia Sinica Institute of Astronomy and Astrophysics and is funded by the Smithsonian Institution and the Academia Sinica. The JCMT was operated by the Joint Astronomy Centre on behalf of the Science and Technology Facilities Council of the UK, the Netherlands Organisation for Scientific Research, and the National Research Council of Canada. Funding for ongoing CARMA development and operations was supported by the NSF and CARMA partner universities.

$Facility$: \facility{EHT}


\begin{thebibliography}{}
\bibitem[Akiyama et al.(2015)]{2015ApJ...807..150A} Akiyama, K., Lu, R.-S., 
Fish, V.~L., et al.\ 2015, \apj, 807, 150 
\bibitem[Alef \& Porcas(1986)]{1986A&A...168..365A} Alef, W., \& Porcas, R.~W.\ 1986, \aap, 168, 365 
\bibitem[Bardeen(1973)]{1973blho.conf..215B} Bardeen, J.~M.\ 1973, Black Holes,
ed. C. DeWitt \& B. S. DeWitt (New York: Gordon and Breach), 215
\bibitem[Benkevitch et al.(2016)]{2016arXiv160900055B} Benkevitch, L., Akiyama, K., Lu, R., Doeleman, S., \& Fish, V.\ 2016, arXiv:1609.00055 
\bibitem[Boehle et al.(2016)]{2016ApJ...830...17B} Boehle, A., Ghez, A.~M., Sch{\"o}del, R., et al.\ 2016, \apj, 830, 17 
\bibitem[Bower et al.(2006)]{2006ApJ...648L.127B} Bower, G.~C., Goss, W.~M.,
Falcke, H., Backer, D.~C., \& Lithwick, Y.\ 2006, ApJL, 648, L127 
\bibitem[Bower et al.(2004)]{2004Sci...304..704B} Bower, G.~C., Falcke, H., 
Herrnstein, R.~M., et al.\ 2004, Science, 304, 704 
\bibitem[Bower et al.(2014)]{2014ApJ...790....1B} Bower, G.~C., Markoff, 
S., Brunthaler, A., et al.\ 2014, \apj, 790, 1 
\bibitem[Bower et al.(2015)]{2015ApJ...798..120B} Bower, G.~C., Deller, A., Demorest, P., et al.\ 2015, \apj, 798, 120 
\bibitem[Broderick \& Loeb(2006)]{2006MNRAS.367..905B} Broderick, A.~E., \& Loeb, A.\ 2006, \mnras, 367, 905
\bibitem[Broderick et al.(2009)]{2009ApJ...697...45B} Broderick, A.~E., Fish,
V.~L., Doeleman, S.~S., \& Loeb, A.\ 2009, \apj, 697, 45 
\bibitem[Chan et al.(2015)]{2015ApJ...799....1C} Chan, C.-K., Psaltis, D., {\"O}zel, F., Narayan, R., \& S{\c a}dowski, A.\ 2015, \apj, 799, 1 
\bibitem[Cotton(1995)]{1995ASPC...82..189C}Cotton, W.~D. 1995, Very Long Baseline Interferometry and the VLBA (ASP Conf. Ser. 82), ed. J. A. Zensus, P. J. Diamond, \& P. J. Napier (San Francisco, CA: ASP), 189
\bibitem[Deller et al.(2011)]{2011PASP..123..275D} Deller, A.~T., Brisken, 
W.~F., Phillips, C.~J., et al.\ 2011, \pasp, 123, 275 
\bibitem[Dexter et al.(2009)]{Dexter2009} Dexter, J., Agol, E., \& Fragile, P.~C.\ 2009, \apjl, 703, L142 
\bibitem[Dexter et al.(2010)]{Dexter2010} Dexter, J., Agol, E., Fragile, P.~C., \& McKinney, J.~C.\ 2010, \apj, 717, 1092 
\bibitem[Dexter \& Fragile(2013)]{2013MNRAS.432.2252D} Dexter, J., \& Fragile, P.~C.\ 2013, \mnras, 432, 2252 
\bibitem[Doeleman et al.(2008)]{2008Natur.455...78D} Doeleman, S.~S., Weintroub,
J., Rogers, A.~E.~E., et al.\ 2008, \nat, 455, 78
\bibitem[Doeleman et al.(2009)]{2009astro2010S..68D} Doeleman, S., Agol, E., Backer, D., et al.\ 2009, astro2010: The Astronomy and Astrophysics Decadal Survey, 2010, 68
\bibitem[Doeleman et al.(2012)]{2012Sci...338..355D} Doeleman, S.~S., Fish,
V.~L., Schenck, D.~E., et al.\ 2012, Science, 338, 355
\bibitem[Eckart et al.(2012)]{2012A&A...537A..52E} Eckart, A., Garc{\'{\i}}a-Mar{\'{\i}}n, M., Vogel, S.~N., et al.\ 2012, \aap, 537, A52 
\bibitem[Falcke \& Markoff(2013)]{2013CQGra..30x4003F} Falcke, H., \& Markoff,
S.~B.\ 2013, Classical and Quantum Gravity, 30, 244003 
\bibitem[Falcke et al.(2000)]{2000ApJ...528L..13F} Falcke, H., Melia, F., \& Agol, E.\ 2000, \apjl, 528, L13  
\bibitem[Fish et al.(2011)]{2011ApJ...727L..36F} Fish, V.~L., Doeleman, S.~S., Beaudoin, C., et al.\ 2011, \apjl, 727, L36 
\bibitem[Fish et al.(2014)]{2014ApJ...795..134F} Fish, V.~L., Johnson, M.~D.,
Lu, R.-S., et al.\ 2014, \apj, 795, 134
\bibitem[Fish et al.(2016)]{2016ApJ...820...90F} Fish, V.~L., Johnson, M.~D., Doeleman, S.~S., et al.\ 2016, \apj, 820, 90
\bibitem[Genzel et al.(2010)]{2010RvMP...82.3121G} Genzel, R., Eisenhauer, F.,
\& Gillessen, S.\ 2010, Reviews of Modern Physics, 82, 3121
\bibitem[G{\"u}sten et al.(2006)]{2006A&A...454L..13G} G{\"u}sten, R., Nyman, L.~{\AA}., Schilke, P., et al.\ 2006, \aap, 454, L13 
\bibitem[Gwinn et al.(2014)]{2014ApJ...794L..14G} Gwinn, C.~R., Kovalev, Y.~Y., Johnson, M.~D., \& Soglasnov, V.~A.\ 2014, \apjl, 794, L14 
\bibitem[Hukushima \& Nemoto(1996)]{1996JPSJ...65.1604H} Hukushima, K., \& Nemoto, K.\ 1996, Journal of the Physical Society of Japan, 65, 1604 
\bibitem[Johnson et al. (2015)]{J2015}Johnson, M.~D., Fish, V.~L., Doeleman, S.~S., et al.\ 2015, Science, 350, 1242 
\bibitem[Johnson \& Gwinn(2015)]{2015ApJ...805..180J} Johnson, M.~D., \& Gwinn, C.~R.\ 2015, \apj, 805, 180 
\bibitem[Johnson \& Narayan(2016)]{2016ApJ...826..170J} Johnson, M.~D., \& Narayan, R.\ 2016, \apj, 826, 170 
\bibitem[Johnson(2016)]{2016ApJ...833...74J} Johnson, M.~D.\ 2016, \apj, 833, 74 
\bibitem[Kamruddin 
\& Dexter(2013)]{2013MNRAS.434..765K} Kamruddin, A.~B., \& Dexter, J.\ 2013, \mnras, 434, 765 
\bibitem[Kawaguchi et al.(2015)]{2015PASJ..tmp..252K} Kawaguchi, N., Jiang, W., \& Shen, Z.-Q. 2015, \pasj, 67, 112 
\bibitem[Kim et al.(2016)]{2016ApJ...832..156K} Kim, J., Marrone, D.~P., Chan, C.-K., et al.\ 2016, \apj, 832, 156 
\bibitem[Kormendy 
\& Richstone(1995)]{1995ARA&A..33..581K} Kormendy, J., \& Richstone, D.\ 1995, \araa, 33, 581
\bibitem[Liddle(2007)]{2007MNRAS.377L..74L} Liddle, A.~R.\ 2007, \mnras, 377, L74
\bibitem[Lu et al.(2012)]{2012ApJ...757L..14L} Lu, R.-S., Fish, V.~L., 
Weintroub, J., et al.\ 2012, \apjl, 757, L14 
\bibitem[Lu et al.(2013)]{2013ApJ...772...13L} Lu, R.-S., Fish, V.~L., 
Akiyama, K., et al.\ 2013, \apj, 772, 13  
\bibitem[Lu et al.(2016)]{2016ApJ...817..173L} Lu, R.-S., Roelofs, F., Fish, V.~L., et al.\ 2016, \apj, 817, 173 
\bibitem[Luminet(1979)]{1979A&A....75..228L} Luminet, J.-P.\ 1979, \aap, 75, 228 
\bibitem[McKinney \& Blandford(2009)]{McKinney2009} McKinney, J.~C., \& Blandford, R.~D.\ 2009, \mnras, 394, L126 
\bibitem[Medeiros et al.(2016)]{2016arXiv160106799M} Medeiros, L., Chan, C.-k., Ozel, F., et al.\ 2016, arXiv:1601.06799 
\bibitem[Melia \& Falcke(2001)]{2001ARA&A..39..309M} Melia, F., \& Falcke, H.\ 2001, \araa, 39, 309
\bibitem[Mo{\'s}cibrodzka et al.(2009)]{Moscibrodzka2009} Mo{\'s}cibrodzka, M., Gammie, C.~F., Dolence, J.~C., Shiokawa, H., \& Leung, P.~K.\ 2009, \apj, 706, 497 
\bibitem[Mo{\'s}cibrodzka 
\& Falcke(2013)]{2013A&A...559L...3M} Mo{\'s}cibrodzka, M., \& Falcke, H.\ 2013, \aap, 559, L3 
\bibitem[Mo{\'s}cibrodzka et al.(2014)]{2014A&A...570A...7M} Mo{\'s}cibrodzka, M., Falcke, H., Shiokawa, H., \& Gammie, C.~F.\ 2014, \aap, 570, A7 
\bibitem[Narayan et al.(2012)]{Narayan2012} Narayan, R., S{\"A} dowski, A., Penna, R.~F., \& Kulkarni, A.~K.\ 2012, \mnras, 426, 3241 
\bibitem[Pearson(1991)]{Pearson1991} T. Pearson, Introduction to the Caltech VLBI Programs, California Institute of Technology, Pasadena, CA, 1991. \seqsplit{http://www.astro.caltech.edu/\textasciitilde{}tjp/citvlb/} 
\bibitem[Rauch et al.(2016)]{2016A&A...587A..37R} Rauch, C., Ros, E., Krichbaum, T.~P., et al.\ 2016, \aap, 587, A37 
\bibitem[Rees(1984)]{1984ARA&A..22..471R} Rees, M.~J.\ 1984, \araa, 22, 471 
\bibitem[Roy et al.(2012)]{R12}Roy, A., Wagner, J., Wunderlich, M., et al.\ 2012, in 11th European VLBI Network (EVN) Symposium and the EVN Users Meeting, ed. P. Charlot, G. Bourda, \& A. Collioud (Trieste: SISSA), 57
\bibitem[Rogers et al.(1995)]{1995AJ....109.1391R} Rogers, A.~E.~E., 
Doeleman, S.~S., \& Moran, J.~M.\ 1995, \aj, 109, 1391 
\bibitem[Shen et al.(2005)]{2005Natur.438...62S} Shen, Z.-Q., Lo, K.~Y., 
Liang, M.-C., Ho, P.~T.~P., \& Zhao, J.-H.\ 2005, \nat, 438, 62 
\bibitem[Wagner et al.(2015)]{2015A&A...581A..32W} Wagner, J., Roy, A.~L., Krichbaum, T.~P., et al.\ 2015, \aap, 581, A32 
\bibitem[Whitney et al.(2004)]{2004RaSc...39.1007W} Whitney, A.~R., 
Cappallo, R., Aldrich, W., et al.\ 2004, Radio Science, 39, RS1007 
\end{thebibliography}
\end{document}